# Laser Assisted Solution Synthesis of High Performance Graphene Supported Electrocatalysts

*Yudong Peng, Jianyun Cao\*, Jie Yang, Wenji Yang, Chao Zhang, Xiaohong Li, Robert A.W. Dryfe, Lin Li, Ian A. Kinloch\*, Zhu Liu\**

Y. Peng, Dr. J. Cao, W. Yang, C. Zhang, Prof. I. A. Kinloch, Dr. Z. Liu
Department of Materials, School of Natural Sciences, University of Manchester, Oxford Road, Manchester, M13 9PL, U.K.
E-mail: jianyun.cao@manchester.ac.uk; ian.kinloch@manchester.ac.uk; zhu.liu@manchester.ac.uk

J. Yang, Prof. R.A.W. Dryfe
Department of Chemistry, School of Natural Sciences, University of Manchester, Oxford Road, Manchester, M13 9PL, U.K.

Dr. J. Cao, J. Yang, Prof. R.A.W. Dryfe, Prof. I. A. Kinloch
National Graphene Institute, University of Manchester, Oxford Road, Manchester, M13 9PL, U.K.

Prof. X. Li
Renewable Energy Group, College of Engineering, Mathematics and Physical Sciences, University of Exeter, Penryn Campus, Cornwall TR10 9FE, U.K.

Prof. L. Li
Laser Processing Research Centre, Department of Mechanical, Aerospace and Civil Engineering, Faculty of Science and Engineering, University of Manchester, Oxford Road, Manchester, M13 9PL, U.K.

**Abstract**

Simple, yet versatile, methods to functionalize graphene flakes with metal (oxide) nanoparticles are in demand, particularly for the development of advanced catalysts. Herein, based on light-induced electrochemistry, we report a laser-assisted, continuous, solution route for the simultaneous reduction and modification of graphene oxide with catalytic nanoparticles. Electrochemical graphene oxide (EGO) was used as starting material and electron-hole pair source due to its low degree of oxidation, which imparts structural integrity and an ability to withstand photo-degradation. Simply illuminating a solution stream containing EGO and metal salt (e.g. $H_2PtCl_6$ or $RuCl_3$) with a 248 nm wavelength laser produced reduced EGO (rEGO, oxygen content 4.0 at.%) flakes, decorated with Pt (~2.0 nm) or $RuO_2$ (~2.8 nm) nanoparticles. The $RuO_2$-rEGO flakes exhibited superior catalytic activity for the oxygen evolution reaction, requiring a small overpotential of 225 mV to reach a current density of 10 mA cm$^{-2}$. The Pt-rEGO flakes (10.2 wt.% of Pt) shows enhanced mass activity for the hydrogen evolution reaction, and similar performance for oxygen reduction reaction compared to a commercial 20 wt.% Pt/C catalyst. This simple production method was also used to deposit PtPd alloy and $MnO_x$ nanoparticles on rEGO, demonstrating its versatility in synthesizing functional nanoparticle-modified graphene materials.

# 1. Introduction

The urgent need for sustainable and clean energy to reduce the usage of traditional fossil fuels has promoted enormous interest in the field of energy storage and conversion.[1] The use of hydrogen as an intermediate for energy storage and power generation has been considered as one of the most promising alternatives to the current non-renewable fossil fuels. Within the hydrogen economy, molecular hydrogen links power grids to other energy sectors through a zero-emission electrochemical pathway.[2-3] In detail, the pathway is the generation of hydrogen as well as oxygen via electrochemical water splitting by the hydrogen evolution reaction (HER) and oxygen evolution reaction (OER) at the generator and then the consumption of hydrogen using a fuel cell system, in which the oxygen reduction reaction (ORR) and hydrogen oxidation reactions convert hydrogen directly into electricity. However, the sluggish kinetics of these electrochemical energy conversion reactions limits seriously the wide application of hydrogen energy.[4] The development of high-performance electrocatalysts, which can reach a designated current density under minimum overpotential, is desirable for maximizing the hydrogen production and utilization efficiency. To-date, platinum group metals and their oxides remain the state-of-the-art catalysts for electrochemical energy conversion.[2] To reduce the cost of electrocatalysts, carbon nanomaterials (e.g. carbon black, carbon nanotubes, graphene, etc.) are widely used as supports for the platinum group metal (oxide) nanoparticles.

Due to its high electrical conductivity and large specific surface area, the two-dimensional (2D) single atom thick graphene has been considered as a promising supporting material for developing advanced electrocatalysts.[5] Compared with the hydrophobic pristine graphene flakes, graphene oxide (GO) with oxygen groups and thus aqueous solution processability has become a more versatile starting material for loading/supporting functional nanoparticles, including electrocatalyst nanoparticles.[6] In the last decade, various methods have been developed to deposit nanoparticles onto the surface of graphene/GO/reduced GO (rGO) flakes, including wet chemical deposition,[7-10] electrochemical deposition,[11-12] and plasma-assisted synthesis,[13] etc. However, several of these techniques require elevated temperatures, harsh chemicals, or high voltage bias. A robust and versatile method which is able to synthesis various types of ultrafine nanoparticles on the surface of graphene in a single-step is still desired.

The use of laser technology to prepare nanomaterials has recently attracted increasing attention due to its simple and fast merits.[14-20] Methods including laser ablation in liquid (PLAL),[14-15] laser pyrolysis,[16, 18, 20-21] and photodeposition[22-24] have been demonstrated for the synthesis of either carbon supported or unsupported nanoparticles for the applications in electrochemical energy storage and conversion. Briefly, in laser ablation in liquid, nanoparticles are formed by the rapid cooling of a plasma plume compromised of elements from the solid ablation targets and the surrounding liquid;[14-15] while in laser pyrolysis, the laser induced carbonization/graphitization of a precursor (e.g. polyimide) and/or decomposition of metal salt precursor leads to the formation of nanoparticles modified graphene/nanocarbon materials.[16, 18, 20-21] Different from the photothermal mechanism of both the laser ablation in liquid and laser pyrolysis, which requires focused and intense laser beam, photodeposition is based on mild light-induced electrochemistry.

Photodeposition of metal (oxide) nanoparticles on the surfaces of semiconductors (metal oxides or sulfides) has been thoroughly studied in the past few decades as indicated in a review paper published recently.[25] Photodeposition is driven by light-induced electron transfer, it occurs simply via illumination of dispersions of semiconductor particles in aqueous solutions containing metal salt precursors.[25] The photoexcitation of semiconductor creates electron-hole pairs, which reduce/oxidize the adsorbed metal ions into metal/metal oxide; the insoluble metal/metal oxide heterogeneously nucleates and grows on the semiconductor substrate.[25-26] In spite of the numerous interests in photodeposition of nanoparticles on the surfaces of metal oxides and sulfides semiconductors, only a few works have studied GO,[22-24] which also behaves as a semiconductor with tunable bandgap values depending on the content and type of oxygen groups.[27-29] Additionally, the sizes of the photodeposited nanoparticles on rGO reported in these pioneering works using lasers with wavelengths of 355 and 532 nm are relatively large (> 5 nm) and not uniformly distributed.[22-24] To date, there is no report regarding the use of photodeposited nanoparticles on GO/rGO as electrocatalysts. One of the possible reasons for the absence of reports on the use of GO for photodeposition is the simultaneous reduction and degradation of GO when exposed to laser irradiation. The reduction of GO by the photon-excited electrons,[25, 30-31] is accompanied by the undesired oxidative GO degradation by photo-generated holes.[31-32] This generally leads to a partial reduction or even degradation of the GO flakes,[31-33] particularly for the heavily oxidized GO with a large amount of oxygen groups (oxygen composition ≥ 30 at.%).[32] In addition, the existence of electron scavenging metal ions (e.g. $Pd^{2+}$) during the photodeposition of metal nanoparticles could further impede the full reduction of GO.[24]

A very recent discovery suggests that the use of mildly oxidized, oxygen functionalized graphene as starting material leads to highly reduced high quality graphene flakes via an ultraviolet (UV) light-induced reduction.[31] Compared with the conventional chemical GO (CGO) produced by Hummers' method, the oxygen functionalized graphene has lower oxygen content and a less disrupted graphene honeycomb lattice structure.[34-35] Electrochemical GO (EGO), produced using a scalable, low-cost and environmentally friendly electrochemical oxidation, has a similar structure with oxygen functionalized graphene with a low content of oxygen groups (~20 at.%), especially the unrestorable C=O and −COO− groups.[36] It has been proved that the reduction of EGO via chemical approaches (e.g. hydrazine) can lead to a higher degree of graphene lattice restoration compared to that can be achieved using CGO.[36] In addition, the photo-degradation of carbon lattice of GO is reported to be dependent on the oxidation level/oxygen content, with a more severe degradation for GO has higher oxygen content.[32] Therefore, the higher structural stability and narrower bandgap of EGO due to the lower oxygen content compared with CGO could potentially allow a rapid and full reduction using lasers with high photon energy and fluence (i.e. laser energy density in mJ cm$^{-2}$) without causing significant degradation of the carbon lattice. Meanwhile, the use of laser beam with higher photon energy and fluence benefits the generation of electron-hole pairs via a one photon process, the abundant electrons/holes would potentially lead to a high nucleation rate for the metal/metal oxide particles and thus small particle size.

Herein, in this work, we report a UV (248 nm) laser-induced continuous solution phase strategy for simultaneous reduction and modification of EGO with uniformly distributed ultrafine catalyst nanoparticles. In a typical experiment (**Figure 1 and S1**), the aqueous precursor solution of metal salt (e.g. $H_2PtCl_6$, $RuCl_3$, etc.) and EGO was circulated continuously from a bulk solution tank to a quartz cell reactor, on which a UV laser (KrF excimer; wavelength: 248 nm; pulse width: 10 ns; repetition rate: 100 Hz; photon energy: 4.99 eV) with a beam size of 1.1 × 0.4 cm$^2$ and irradiates at various laser fluences. This laser-induced solution approach leads to deeply reduced EGO (rEGO) flakes with a low oxygen content of 4.0 at.% and a partial restoration of the graphene lattice structure evidenced by Raman spectroscopy. The as-formed Pt and $RuO_2$ nanoparticles distribute uniformly on the rEGO flakes, with ultrafine average diameters of 2.0 nm (σ = 0.5) and 2.8 nm (σ = 0.6), respectively. When used as electrocatalysts, the $RuO_2$-rEGO exhibits superior activity for the OER, with a small overpotential of 225 mV at a current density of 10 mA cm$^{-2}$, outperforming the majority of the reported Ru based electrocatalysts. In comparison with the commercial (CM) Pt/C catalyst (Pt loading: 20 wt.%; HISPEC 3000; Johnson Matthey), the as-synthesized Pt-rEGO catalyst with a Pt loading of 10.2 wt.% shows significantly enhanced mass activity for HER, together with comparable performance for ORR. This simple, scalable and straightforward method has been further applied to the synthesis of other metal (oxide) nanoparticles supported on rEGO flakes, including PtPd alloy and $MnO_x$, demonstrating its versatility in the production of functional nanoparticle modified graphene materials.

## 2. Results and discussion

### 2.1. Characterization of reduced EGO

The optical bandgap of EGO and their fragments dispersed in water has been estimated by applying the Tauc plot[37] through the UV-Vis absorption spectrum (**Figure 2a and b**). The combination of the π-state ($sp^2$ bonded) carbons, and the σ-state ($sp^3$ bonded) carbons in GO makes it a semiconductor with a bandgap in a range from 2 to 7 eV.[28] The absorption at ~4 eV caused by n-π* transitions of C=O, and the peak at approximately 5 eV is attributed to π-π* transitions of C=C.[38-39] The bandgap values of EGO has been approximated from the linear extrapolation using the Tauc plot,[40] which gives a direct bandgap range from 3.25 to 3.95 eV (**Figure 2a**), and an indirect bandgap from 2.04-2.4 eV (**Figure 2b**). Therefore, the photon energy of 248 nm laser (4.99 eV) is sufficiently high to excite EGO and thus create electron-hole pairs. In contrast, due to the higher degree of oxidation compared with EGO,[36] CGO exhibits larger bandgap values for both direct (3.25 to 4.31 eV) and indirect (2.04 to 3.38 eV) bandgaps **(Figure S2)**.

Raman spectroscopy was used to characterize the quality of the graphene lattice and defect density of rEGO. The excitation laser (He-Ne; 633 nm) output power for recording the Raman spectra was limited to < 0.5 mW to avoid any laser-induced structure alteration. **Figure 2c** shows typical Raman spectra of EGO and rEGO. Spectra of all samples show a D band at 1333 cm$^{-1}$ representing the edge planes and disordered structures, and the characteristic G band at 1595 cm$^{-1}$ ascribed to the ordered sp$^2$ bonded carbon.[41] The Raman spectra were further analyzed by fitting the D and G bands with Lorentzian function after baseline subtraction (see details in **Figure S3**). The intensity ratio of D to G band ($I_D/I_G$) is closely related to the density of defect/functionality in the graphene lattice,[42-44] the $I_D/I_G$ ratio of rEGO increases linearly from 1.25 to 2.01 with the increasing of laser fluence (0 to 681 mJ cm$^{-2}$, Note: zero fluence represents pristine EGO) (**Figure 2d**). Using the model proposed by Lucchese et al.[42] and Cançado et al.,[43] the defect distance ($L_D$) can be determined from the Raman $I_D/I_G$ ratio (**Figure S4a** and **Table S1**). With the increasing laser fluence, $L_D$ rises from 1.17 to 1.32 nm, suggesting a partial restoration of the graphene lattice. Quantification of defect density ($\theta$), defined as the ratio of C(sp$^3$) to C(sp$^2$), using the calculated $L_D$ values,[44] suggests a gradual reduction of $\theta$ from 2.13% to 1.68% with the increase of laser fluences (**Figure S4b** and **Table S1**). The lattice restoration after laser reduction is also confirmed by the narrowing of full width at half-maximum (FWHM; Γ) of D and G bands (**Figure 2e and f**).[45] In addition, the photographs of EGO solutions (**Figure 2g**) indicate that the color of EGO dispersions turned from brown into black after laser irradiation at various fluences of 227, 340, 454, 568 and 681 mJ cm$^{-2}$, respectively. X-ray diffraction (XRD) further confirms the increased reduction of EGO at higher laser fluence. The diffraction peak of EGO at ~10° vanished gradually with the increase of laser fluence (**Figure S5**).

The X-ray photoelectron spectroscopy (XPS) survey spectra collected with pristine EGO and rEGO after irradiation at a laser energy density of 568 mJ cm$^{-2}$ reveal a significant decrease in oxygen content, from 24.0 at.% (EGO) to 4.0 at.% (rEGO) (**Figure S6**). **Figure 2h** compares the C 1s spectra for pristine EGO and rEGO. The deconvolution of the C 1s spectrum of pristine EGO shows five distinctive components, which are assigned to sp$^2$ carbon (284.2 eV), sp$^3$ carbon (284.9 eV), C−O (286.0 eV), C−O−C (287.0 eV) and C=O/−COO− (288.2 eV),[46] (details in **Table S2**). In contrast, the C 1s of rEGO shows an effective removal of oxygen groups and restoration of $sp^2$ carbon structure. Notably, the $sp^2$ carbon content increases dramatically from 2.5 at. % for pristine EGO to 61.9 at.% for rEGO. Both the Raman and XPS results indicate an efficient restoration of the $sp^2$ bonded graphene lattice from the $sp^3$ oxygenated sites via the facile laser-induced reduction. In comparison, the reduction of EGO by thermal annealing up to 250 °C removes oxygen functionalities but leaves the defects unrestored. The XRD diffraction peak of EGO at ~ 10° decreases in its intensity and disappears after annealing at temperature > 200 °C, revealing the removal of the majority of the oxygen functionalities (**Figure S7**). Previous reports also suggested that thermal annealing at 200 °C effectively reduced EGO film as evidenced by the dramatic recovery of electrical conductivity.[36] However, the decrease of Raman $I_D/I_G$ ratio (**Figure S8**) for thermal annealed EGO suggests the $sp^2$ bonded graphene lattice structure is not restored. Quantification and comparison of defect distance and density using Raman $I_D/I_G$ ratio indicate that thermal annealing leads to a slightly increased defect density compared with pristine EGO (**Figure S9**). This is due to the thermal decomposition of oxygen functional groups to $CO_2$/CO products, thereby removing oxygen and carbon atoms simultaneously, leaving permanent defects and vacancies in the graphene lattice.[47] The results from thermally reduced EGO also suggest that the UV laser-induced reduction of EGO in aqueous solution follows a dominating photochemical mechanism with minimum photothermal effect.

Due to the hydrophobic nature of graphene, the rEGO dispersion agglomerated and settled at the bottom of the water within 2 hours (**Figure S10**). Additionally, there is a gradual increase in weight loss (26.8 to 79.1 wt.%) and Raman $I_D/I_G$ ratio (1.46 to 2.01) for rEGO irradiated using a laser beam with increasing fluence from 227 to 568 mJ cm$^{-2}$ (**Figure S11**). This suggests a more efficient removal of oxygen groups and restoration of graphene lattice at higher laser fluence. Further increase of the laser fluence to 681 mJ cm$^{-2}$ causes a sudden drop in the remaining weight of rEGO to 22.4% compared to the starting EGO (**Figure S11**), indicating photodegradation occurs at such high laser fluence. In comparison with EGO, there are much less apparent changes in weight loss, Raman $I_D/I_G$ ratio, $\Gamma_D$ and $\Gamma_G$ for CGO after laser irradiation at increasing fluence from 227 to 681 mJ cm$^{-2}$ (**Figure S12**), but a more severe photodegradation (13.8% remaining weight) at the laser fluence of 681 mJ cm$^{-2}$. In comparison to the rEGO, the laser treated CGO under the same condition yields a brownish suspension indicating insufficient reduction (**Figure S13**). The less efficient reduction and restoration of conventional CGO using the laser-assisted solution approach are consistent with the recent report.[31] To avoid severe photodegradation, the laser fluence of 568 mJ cm$^{-2}$ was selected to be optimal for the simultaneous reduction of EGO and deposition of catalyst nanoparticles.

## 2.2. Characterization of reduced EGO modified with metal (oxide) nanoparticles

Metal salts of chloroplatinic acid ($H_2PtCl_6$) and ruthenium chloride ($RuCl_3$) were selected as precursors for metal (Pt) and metal oxide ($RuO_2$) nanoparticles, respectively. **Figure 2i** shows the Raman spectra of $RuO_2$-rEGO and Pt-rEGO in comparison with pure rEGO reduced by UV laser irradiation. All samples exhibit sharp D and G bands with broad 2D and D + D` bands. The $I_D/I_G$ ratios for $RuO_2$-rEGO, Pt-rEGO and rEGO are 1.71, 2.05 and 1.83, respectively, which are significantly different from the ratio of the original EGO ($I_D/I_G$ = 1.25), indicating the laser-induced reduction of EGO remains effective after the addition of metal salts in the dispersion. Interestingly, the $RuO_2$-rEGO sample shows a reduced $I_D/I_G$ ratio compared with that of rEGO, suggesting more defects/functionalities. This indicates a possible bond formation (oxygen bridges) between rEGO and $RuO_2$. In addition, the G band position of $RuO_2$-rEGO sample upshifts by 5.7 cm$^{-1}$ to 1597.8 cm$^{-1}$ (**Figure S14**) compared with that of rEGO (1592.1 cm$^{-1}$), indicating charge transfer occurred between rEGO and $RuO_2$ via doping effect and/or bond formation.[48] Therefore, the reduced $I_D/I_G$ ratio and shift of G band position of RuO$_2$-rEGO compared with rEGO suggests the formation of oxygen bridges. The complete reduction of EGO in the Pt-rEGO was confirmed by XPS C 1s spectrum (**Figure S15a**). In contrast, according to literature,[24] the photodeposition of Pd nanoparticles from Pd$^{2+}$ inhibited the complete reduction of CGO. As the reduction potential vs standard hydrogen electrode (SHE) for PdCl$_4^{2-}$/Pd (0.591 V) is not significantly different from that of PtCl$_6^{2-}$/PtCl$_4^{2-}$ (0.68 V) and PtCl$_4^{2-}$/Pt (0.755 V).[49] The effective laser-induced reduction of EGO in the presence of competing reactions (reduction of PtCl$_6^{2-}$ to Pt) suggests EGO with low contents of oxygen (~20 at.%) and unrestorable C=O/−COO− groups as a promising platform for the laser-assisted synthesizing of nanoparticle-functionalized graphene materials.

Both the Raman spectrum and XRD pattern (**Figure S16a and b**) of the as-synthesized RuO$_2$-rEGO indicate an amorphous structure of the freshly deposited hydrous RuO$_2$ (RuO$_2$ with structural water, RuO$_2 \cdot x$H$_2$O). As the degree of crystallinity of RuO$_2$ affects its OER[50-51] and pseudo-capacitance[52-54] performance, heat treatment was performed with the RuO$_2$-rEGO in air at various temperatures ranging from 100 to 250 °C. The Raman spectra of samples annealed at 200 and 250 °C (**Figure S16a**) show characteristic vibrational peaks at 515, 630 and 703 cm$^{-1}$, corresponding to E$_g$, A$_{1g}$ and B$_{2g}$ modes of crystalline RuO$_2$, respectively. The transition from amorphous to crystalline structure for the RuO$_2$-rEGO composite was further confirmed by XRD (**Figure S16b**). The diffraction peaks at 28.0° and 35.0° corresponding to (110) and (101) planes of rutile RuO$_2$ (ICDD No 00-040-1290) appear after annealing at 200 and 250 °C. Interestingly, apart from the RuO$_2$ peaks, weak peaks at 39.4°, 42.2° and 44.1° corresponding to metallic Ru are also observed, indicating the formation of minor metallic Ru phases with the prevailing RuO$_2$ phase. The XRD pattern of Pt-rEGO (**Figure S15b**) shows that all of the diffraction peaks match with Pt (ICDD No. 00-004-0802), suggesting that the successful photodeposition of Pt on the rEGO sheets had occurred. According to the standard reduction potentials of aqueous Ru and Pt solutions,[25, 49] the reduction potential vs SHE decreases in the order: $PtCl_4^{2-}$/Pt (0.755 V) > $PtCl_6^{2-}$/$PtCl_4^{2-}$ (0.68 V) > $Ru^{2+}$/Ru (0.455 V) > $Ru^{3+}$/$Ru^{2+}$ (0.249 V). The deposition of metallic Ru requires a more negative potential than that of Pt, which is probably one of the reasons for the formation of $RuO_2$ as the dominant phase rather than metallic Ru. Note that the pH values of EGO-$H_2PtCl_6$ (3.18) and EGO-$RuCl_3$ (3.32) precursor solutions are comparable. At this pH range, according to the Pourbaix diagram, the reduction potential for $Ru(OH)_3$/Ru is ~ 0.5 V vs SHE, lower than that of $Pt(OH)_2$/Pt (~0.8 V vs SHE).[55] In addition, Ru is known to be much less noble than other Pt group metals and thus has a stable oxidation state of +4 in the presence of oxygen.[55] Electrochemical investigation of Ru metal electrodes indicated that the surface oxidation of Ru already begins at the potentials in, or close to, the H region, 0.05 to 0.2 V vs reversible hydrogen electrode (RHE) in 0.5 M $H_2SO_4$.[56] In contrast, the surface oxidation of Pt starts at a high potential of 0.8 V vs RHE in 0.5 м $H_2SO_4$.[57] Hence, another possible reason for the formation of $RuO_2$ as dominating phase is due to the as-formed metallic Ru is prone to be oxidized by photo-generated holes.

Transmission electron microscopy (TEM) and scanning transmission electron microscopy with high angle annular dark field (STEM-HAADF) were used to characterize the particle size and crystal structure of the $RuO_2$-rEGO and Pt-rEGO composites. The TEM and STEM-HAADF images for both the $RuO_2$-rEGO-250HT (heat-treated in air at 250 °C; **Figure 3a and c**) and Pt-rEGO (**Figure 3b and d**) show uniformly distributed ultrafine nanoparticles on the rEGO support. Compared with the TEM images of the as-synthesized $RuO_2$-rEGO without heat treatment (**Figure S17**), annealing at 250 °C shows minor effects in the morphology of $RuO_2$-rEGO sample. Statistical particle size analysis was conducted for the Pt and $RuO_2$ nanoparticles, with 1000 and 833 particles in random areas being analyzed, respectively, and a log-normal function being used for data fitting. The results show mono-dispersed particle sizes of 2.0 nm (standard deviation: σ = 0.5) for Pt-rEGO and 2.8 nm (σ = 0.6) for RuO$_2$-rEGO-250HT. One of the possible reasons, that the smaller size of nanoparticles (2~3 nm) in this work compared with the sizes of photodeposited nanoparticles (> 5 nm) on rGO in the previous reports,[22-24] is due to the use of laser with higher photon energy and fluence. Analog to electrodeposition, higher laser photon energy and fluence correspond to larger current density (overpotential) and thus increased supersaturation, leading to higher nucleation rate and reduced critical cluster size, therefore smaller particle size.[58] In addition, the possible bond formation (e.g. oxygen bridges) between rEGO and nanoparticles would provide anchoring effect and thus inhibit the agglomeration and growth of small particles. Nevertheless, factors affecting the nanoparticle size of a laser-induced photodeposition process are complicated, the potential factors include but not limited to: concentrations of EGO and metal ions, type and concentration of sacrificial electron donor/acceptor, temperature, pH, laser wavelength (photon energy), fluence, irradiation time, pulse repetition rate, mass diffusion related variables (e.g. stirring, flow rate), etc. Full understanding of the nanoparticle formation mechanism and precise control of nanoparticle size require future and in-depth studies.

Energy-dispersive X-ray spectroscopy (EDS) was employed to analyze the distribution of Ru and Pt elements in the rEGO support. The EDS mapping data (**Figure 3e to g and i to k, respectively**) indicates that both the Ru and Pt are evenly distributed over the entire rEGO support. The HRTEM image of Pt-rEGO (**Figure 3l**) shows the Pt nanoparticles have a single-crystalline feature with clear lattice fringes. The labelled Pt nanoparticles in **Figure 3l** show lattice spacing values of 0.224 and 0.199 nm, which are indexed to the (111) and (200) facets of Pt, respectively. The nanoparticles on RuO$_2$-rEGO-250HT sample show clear lattice fringes (**Figure 3h**) with a spacing value of 0.227 nm, corresponding to the (200) facet of RuO$_2$. In addition, secondary metallic Ru nanoparticles are also observed from the TEM image of RuO$_2$-rEGO sample (**Figure 3h**). Detailed TEM characterization of RuO$_2$-rEGO samples (**Figure S18a to f**) suggests that the majority of the nanoparticles with a relatively large average size of 10.7 nm (σ = 5.7) are metallic Ru. The observation of secondary metallic Ru phase from TEM characterization is consistent with the XRD results. Scanning electron microscope (SEM) images of the drop casted RuO$_2$-rEGO and Pt-rEGO composite films on silicon wafers indicate a thick porous structure (**Figure S19a, c and e**). At higher magnifications, the SEM images of the composite films (**Figure S19b, d and f**) all exhibit a typical wrinkled flake-like morphology of stacked graphene flakes.

XPS was performed to investigate the oxidation state of the elements in the as-formed nanoparticles, and their coupling with the rEGO support. As shown in **Figure 3m,** the Ru 3d high-resolution spectrum of the as-synthesized RuO$_2$-rEGO without heat treatment shows a set of doublet peaks located at 280.9 and 284.9 eV, corresponding to the doublet peaks for Ru (IV) 3d$_{5/2}$ and 3d$_{3/2}$, respectively.[59] Owing to the strong interference of Ru 3d and C 1s signals, the comparison of survey, Ru 4d and Ru 3p spectra (**Figure S20a, b and c, respectively**) for RuO$_2$-rEGO, commercial RuO$_2$ (CM RuO$_2$) and metallic Ru was conducted to identify the oxidation state of Ru. In comparison to the Ru 3p$_{3/2}$ peaks of CM RuO$_2$ at 462.5 eV and metallic Ru at 461.5 eV (**Figure S20c**), the Ru 3p$_{3/2}$ of RuO$_2$-rEGO contains only Ru (IV) with Ru 3p$_{3/2}$ peak located at 462.7 eV. This indicates that the RuO$_2$-rEGO is dominated by Ru (IV) with a negligible amount of Ru (0) (details in **Table S3**). The comparison of Ru 3d (**Figure S20b**) also indicates the dominance of Ru (IV) in RuO$_2$-rEGO. In addition, although the C 1s signal is overlapped with that of Ru 3d$_{3/2}$, three components can be deconvoluted as shown in **Figure 3m**, namely, C−C at 284.6 eV, C−O at 285.6 eV and C=O at 288.6 eV. The existence of C−O bond is possibly due to the strong coupling between rEGO and RuO$_2$ via oxygen bridges instead of insufficient reduction of EGO. The evidence for the formation oxygen bridges includes: (1) the relative lower $I_D/I_G$ ratio and the upshift of G band position for RuO$_2$-rEGO compared with the rEGO indicate possible bond formation via oxygen bridges; (2) deconvolution of overlapped Ru 3d$_{3/2}$ and C 1s spectra of RuO$_2$-rEGO shows that percentage of C−O component in total C 1s spectrum is 19.8 at.%, much higher than that of rEGO (8.8 at.%, Table S2), this high percentage of C−O component in RuO$_2$-rEGO is thereby unlikely from the unreduced functional groups but oxygen bridges between RuO$_2$ and rEGO; (2) the component corresponding to C−O in the O 1s spectrum of RuO$_2$-rEGO is shifted to higher energy compared with that of EGO (**Figure S20d**), suggesting a lower electron density at the oxygen sites due to the electron transfer from the oxygen to Ru atoms;[60] (3) the peak identified at 528.6 eV in O 1s spectrum is in agreement with the bridged O connecting Ru and C as-reported in the literature.[61] For the Pt 4f high-resolution spectrum of Pt-rEGO (**Figure 3n**), the set of doublet peaks located at 71.2 and 74.6 eV are ascribed to the surface Pt atoms, while the doublet peaks at higher binding energies of 72.3 and 75.6 eV correspond to the bulk atoms of Pt.[62] Further, inductively coupled plasma-optical emission spectrometry (ICP-OES) reveals that the mass loadings of RuO$_2$ and Pt in the as-prepared composites are 41.6 ± 0.9 wt.% and 10.2 ± 1.0 wt.%, respectively.

The laser-induced solution approach leads to deeply reduced rEGO modified with ultrafine catalyst nanoparticles in a single step, as demonstrated above using RuO$_2$ and Pt as model materials. To further prove the versatility of this laser-assisted approach in deposition of various types of functional nanoparticles on the surface of rEGO sheets, PtPd alloy and MnO$_x$ nanoparticles have also been successfully deposited using H$_2$PtCl$_6$/Na$_2$PdCl$_4$ and MnCl$_2$ as precursor metal salts, respectively. The corresponding TEM characterizations for PtPd/rEGO and MnO$_x$/rEGO are available in **Figure S21** and **Figure S22**, respectively. Based on the experimental results and literature knowledge,[23, 25, 31] possible reactions behind this laser-induced solution approach have been proposed. The photon excitation of EGO creates electron-hole pairs as:

$$EGO + h\nu \rightarrow EGO + h_{vb}^{+} + e_{cb}^{-} \qquad (1)$$

where $h_{vb}^{+}$ and $e_{cb}^{-}$ are photo-generated holes and electrons, respectively. The subsequent reduction of EGO by photo-generated electrons occurs following:

$$EGO + e_{cb}^{-} \rightarrow rEGO + OH^{-} \qquad (2)$$

For reductive photodeposition of metal (M) nanoparticles, the reduction of metal ions happens as:

$$M^{n+}(aq) + ne_{cb}^{-} \rightarrow M(s) \qquad (3)$$

While for the oxidative photodeposition of metal oxide nanoparticles, the reaction could occur through:

$$M^{n+}(aq) + nh_{vb}^{+} + nH_2O \rightarrow MO_n(s) + 2nH^{+} \qquad (4)$$

The excess photo-generated holes and electrons are consumed by the sacrificial electron donor (D) and acceptor (A), respectively, as follow:

$$nh_{vb}^{+} + D \rightarrow D^{n+} \qquad (5)$$

$$ne_{cb}^{-} + A \rightarrow A^{n-} \qquad (6)$$

In an actual photodeposition process, the consumption of excess electrons leads to the reduction of protons to form $H_2$ gas, while the consumption of excess holes could oxidize water to from $O_2$.[25] If sacrificial organic agents (e.g. methanol, isopropanol) are added, their reactions with photo-generated holes/electrons form highly reducing radical species, which participate in the reduction of metal ions.[25, 31] In the present work, isopropanol and acetone were added as sacrificial electron donor and acceptor, respectively, leading to the formation of highly reducing carbon centered isopropanol radicals,[31] which further benefit a fast and thorough reduction of EGO and metal ions. In addition, these highly reducing radicals could also be the reason that metallic Ru is present in the $RuO_2$-rEGO product.

## 2.3. Electrocatalytic performance

To demonstrate the applications of the as-prepared $RuO_2$-rEGO and Pt-rEGO composites, their performances as electrocatalysts have been measured and evaluated. To obtain the intrinsic catalytic activities of the samples for OER, HER and ORR, the ohmic-drop correction was carried out to minimize the effects of solution resistance (**Figure S23**). For $RuO_2$-rEGO, the OER polarization voltammogram was obtained at a scan rate of 10 mV s$^{-1}$ in $O_2$-saturated 1.0 M KOH aqueous electrolyte using a rotating disk electrode (RDE) at a rotating speed of 2000 rpm. As shown in **Figure 4a**, the $RuO_2$-rEGO-250HT possesses a superior catalytic activity with a small overpotential of 225 mV at 10 mA cm$^{-2}$, compared to the overpotential for the sample heat-treated at 200 °C ($RuO_2$-rEGO-200HT), which is significantly larger (310 mV). There is no obvious catalytic activity for the sample annealed at 150 °C (**Figure S24**). In addition, there is a noticeable drop in the Tafel slope for the $RuO_2$-rEGO composites with the increase of heat treatment temperature from 200 to 250 °C (**Figure 4b**). These results suggest the OER activity is profoundly affected by the crystallinity of $RuO_2$. Note that it is known that metallic Ru nanoparticles are unstable and dissolve completely during the first OER polarization,[63] and this dissolution of Ru is more severe in alkaline electrolytes than in acidic electrolytes.[64] Therefore, the small content of Ru nanoparticles in $RuO_2$-rEGO catalyst would dissolve rapidly in the first OER polarization and have minimum influence on the subsequent evaluation of catalytic activity. This has been confirmed by the first 11 OER CV scans recorded with the $RuO_2$-rEGO catalyst (**Figure S25**), the current due to Ru oxidation/dissolution appears only in the first anodic scan, and the CVs overlap with each other after the second cycle. As a benchmark, the OER performance of CM $RuO_2$ (Premion, Alfa Aesar) was measured under the same conditions. Notably, CM $RuO_2$ requires an overpotential of 283 mV to reach the current density of 10 mA cm$^{-2}$ (**Figure 4a**), with $RuO_2$ loading 2.4 times higher than that of $RuO_2$-rEGO (42.5 wt.% of $RuO_2$). In addition, as shown in **Figure 4b,** the $RuO_2$-rEGO-250HT shows a smaller Tafel slope (50 mV

dec$^{-1}$) than that of CM RuO$_2$ (53 mV dec$^{-1}$). Chronopotentiometric testing at 10 mA cm$^{-2}$ was used to evaluate the durability of RuO$_2$-rEGO-250HT catalyst, as shown in the inset of **Figure 4a**, the overpotential of RuO$_2$-rEGO-250HT increased by only 14.1 mV after 3 hours testing. The comparison of linear sweep voltammograms (LSVs) (**Figure S26a**) and Tafel plots (**Figure S26b**) for the RuO$_2$-rEGO-250HT catalyst before and after chronopotentiometric test shows only slight degradation of catalytic activity.

The electrochemically active surface area (ECSA) of the catalysts has been estimated by measuring the double-layer capacitance as reported in the literature.[65] The $C_{DL}$ was determined from the cyclic voltammetry (CV) scans in the non-Faradaic region between −0.1 and 0.1 V vs Ag/AgCl in 1 M KOH aqueous electrolyte (**Figure S23 c and d**). The ECSA was estimated to be 50.2 and 47.4 cm$^2$ for RuO$_2$-rEGO-250HT and CM RuO$_2$, respectively, indicating that the RuO$_2$-rEGO composite with 42.5 wt.% of RuO$_2$ has a slightly higher accessible surface area to electrolyte in comparison with CM RuO$_2$ (100 wt.% RuO$_2$ loading). In addition, the specific surface areas derived from nitrogen adsorption/desorption isotherms (**Figure S27**) also suggest that the RuO$_2$/rEGO has a higher specific surface area (261.8 m$^2$ g$^{-1}$) than that of CM RuO$_2$ (49.1 m$^2$ g$^{-1}$). Electrochemical impedance spectroscopy (EIS) was conducted at 1.5 V vs RHE for the as-prepared catalysts (**Figure S28**). Interestingly, the charge transfer resistances derived from the Nyquist plots for RuO$_2$-rEGO composites (6.72 and 10.59 Ω for RuO$_2$-rEGO-250HT and RuO$_2$-rEGO-200HT, respectively) are significantly smaller than that of CM RuO$_2$ (18.05 Ω). The enhanced charge transfer in the RuO-rEGO composites is attributed to the highly conductive rEGO support, which provides fast electron transfer routes. In addition, the smaller charge transfer resistance for RuO$_2$-rEGO composite annealed at 250 °C, compared to that annealed at 200 °C, is probably due to increased electrical conductivity of RuO$_2$ itself at higher annealing temperatures.[66]

As revealed in **Figure 4c**, the RuO$_2$-rEGO-250HT exhibits much higher (one order of magnitude) mass activity and turnover frequency (TOF) than that of the CM RuO$_2$ (A detailed TOF calculation is given in Supporting Information). This further confirms the enhancement of intrinsic catalyst activity for the RuO$_2$-EGO composites. The OER performance of RuO$_2$-rEGO-250HT has been further compared with the state-of-the-art Ru-based catalysts reported in the literature. The overpotential of RuO$_2$-rEGO-250HT at 10 mA cm$^{-2}$ and the Tafel slope are outperforming the majority of the literature values (**Figure 4d** and **Table S4**). The low overpotential, small Tafel slope, high mass activity and TOF, together with the good durability of the RuO$_2$-rEGO composites indicate their great potential as electrocatalysts for OER.

RuO$_2$ is known to be a very promising electrode material for supercapacitors due to its large specific capacitance (experimental values up to 720 F g$^{-1}$) and good rate capability.[66] However, the rarity of Ru in Earth's crust results in the high cost of RuO$_2$, which limits its wider application in the real world. This challenge could be possibly solved by developing composites of RuO$_2$ and carbon, which can reduce the usage of RuO$_2$ while maintaining high specific capacitance. Hence, the electrochemical capacitance of RuO$_2$-rEGO composites (RuO$_2$ loading: 41.6 ± 0.9 wt.%) has been measured and evaluated.

Initially, the effect of annealing temperature on the electrochemical capacitance of the RuO$_2$-rEGO composite was investigated (**Figure S29**). The specific capacitance increases with the rise of annealing temperature and maximizes at 200 °C, further increase of the annealing temperature causes deterioration of the specific capacitance. This phenomenon is consistent with previous works and can be explained by the balancing of ionic (proton), and electronic conductivity of hydrous RuO$_2$ during annealing (detailed discussion in **Figure S29**).[66] **Figure 4e** shows the CVs of RuO$_2$-rEGO composite annealed at 200 °C at a scan rate of 20 mV s$^{-1}$ in 1 M H$_2$SO$_4$ electrolyte. The CV of CM RuO$_2$ recorded under the same conditions is added for comparison. Both the CVs show a typical rectangular shape corresponding to capacitive behavior. A pair of broad redox peaks appears at around 0.4 V vs Ag/AgCl due to the Ru (III)/Ru (IV) transition. **Figure 4f** compares the gravimetric specific capacitances of $RuO_2$-rEGO-200HT composite and the CM $RuO_2$. Note that the specific capacitance of $RuO_2$-rEGO-200HT is normalized by the total mass of the composite. The $RuO_2$-rEGO-200HT shows higher specific capacitance of 649.7 F $g^{-1}$ at a current density of 0.5 A $g^{-1}$ with good rate capability, outperforming the CM $RuO_2$ (518.2 F $g^{-1}$ at 0.5 A $g^{-1}$).

For the Pt nanoparticle modified rEGO (Pt loading 10.2 ± 1.0 wt.%), the catalytic performances with respect to ORR and HER have been measured and compared with CM Pt/C (20 wt.%, HISPEC 3000, Johnson Matthey). **Figure 5a** shows the LSV response of Pt-rEGO catalyst as a function of rotation speed in an $O_2$ saturated 0.1 M KOH aqueous electrolyte. The Koutechy-Levich (K-L) relation of the LSV curves (inset of Figure 5a) exhibits good linearity with an average electron transfer number (*n*) of 3.93, suggesting a four-electron reduction to water is largely operative. In addition, the Pt-rEGO catalyst shows a higher double layer capacitance in comparison with CM Pt/C according to the CVs recorded in $N_2$-saturated electrolyte (**Figure S30**), possibly due to the large surface area of rEGO. The ECSA of the catalysts were further determined through integration of the charge for hydrogen adsorption and desorption in a $N_2$ saturated environment. Due to the ultrafine size of the Pt nanoparticles (2.0 ± 0.5 nm), the Pt-rEGO provides an enhanced ECSA of 97.2 $m^2$ $g_{Pt}^{-1}$ compared with that of CM Pt/C (73.9 $m^2$ $g_{Pt}^{-1}$). As displayed in **Figure 5b**, Pt-rEGO exhibits a similar onset potential (0.95 V), and half-wave potential (0.833 V) to that of the CM Pt/C, indicating comparable activities of Pt-rEGO and CM Pt/C in spite of the much lower Pt loading (10.2 wt.% vs 20 wt.%). Tafel plots (inset of Figure 5b) suggest a slightly enhanced Tafel value of 78.5 mV $dec^{-1}$ for Pt-rEGO compared with that of the Pt/C (81.5 mV $dec^{-1}$), indicating fast kinetics of ORR with Pt-rEGO catalyst. However, the diffusion-limited current density of ORR for Pt-rEGO (4.89 mA cm$^{-2}$) is slightly smaller than that of CM Pt/C (5.36 mA cm$^{-2}$). This is caused by retarded diffusion of dissolved O$_2$ through the stacked graphene sheets.[67] Effective prevention of graphene restacking could lead to further enhancement in performance, but it remains a big challenge. Nevertheless, the 2D rEGO flakes act as barriers to prevent the leaching and dissolution of Pt,[67] leading to significantly improved ORR durability for Pt-rEGO catalyst (76% retention after ca. 5.6 hours test) compared with that of CM CM Pt/C (63% retention after ca. 5.6 hours; see **Figure S31a** for details).

The Pt-rEGO exhibits superior catalytic activity for HER when compared with the CM Pt/C. The HER activity of Pt-rEGO was measured in N$_2$ saturated 0.5 M H$_2$SO$_4$ electrolyte. **Figure 5c** shows the polarization curves of Pt-rEGO and CM Pt/C catalysts after ohmic-drop correction. To drive 10 mA cm$^{-2}$ current density, the Pt-rEGO only required a small overpotential of 28.27 mV which is superior to that of CM Pt/C (33.13 mV). The Tafel plots (inset of **Figure 5c**) show comparable kinetics of Pt-rEGO (32.5 mV dec$^{-1}$) and CM Pt/C (30 mV dec$^{-1}$). Similar to ORR, the barrier effects of 2D rEGO leads to an obviously improved durability of Pt-rEGO for HER compared with that of CM Pt/C (**Figure S31b**). Figure 5d reveals the mass activities of Pt-rEGO catalyst for both HER and ORR, in comparison with that of CM Pt/C. Since the Pt loading Pt-rEGO (10.2 wt.%) is around half the value of CM Pt/C (20 wt.%), it shows more than double the mass activity for HER at overpotentials of 10, 30 and 50 mV. The statistics of the Pt particles in CM Pt/C shows an average size of 4.3 nm ($\sigma$ = 1.0) (**Figure S32**), which is twice the size of Pt particles in Pt-rEGO (2.0 nm $\sigma$=0.5). The results imply that the facilitated catalytic performance of Pt-rEGO can be possibly ascribed to smaller particle size and thus higher utilization of Pt atoms. For ORR, the Pt-rEGO exhibits comparable mass activity with CM Pt/C at both 0.85 and 0.9 V vs RHE, in spite of the slightly lower diffusion-limited current density due to the inhibited oxygen diffusion in the restacked graphene layers.

The RuO$_2$ and Pt modified rEGO show superior catalytic activity for the water splitting processes (OER and HER) compared with the commercial catalysts and literature values, together with comparable ORR activity (see **Table S5-S6** for detail comparison). The better performance of the Pt modified rEGO catalysts for HER compared with ORR is possibly because that the diffusion limitation in the restacked rEGO layers has a lower effect on HER due to the high concentration of H$^+$ in the acidic electrolyte.[68]

## 3. Conclusions

In summary, a simple, yet versatile, UV laser assisted method based on photo-induced redox processes has been demonstrated in this work for continuous solution reduction and modification of EGO with functional nanoparticles. The use of EGO with a lower degree of oxidation, and thus narrower bandgap, and better structural integrity compared with conventional CGO leads to deeply reduced rEGO with a low oxygen content of 4.0 at.%. Various types of ultrafine metal (oxide) nanoparticles (Pt, PtPd, RuO$_2$, MnO$_x$) have been uniformly deposited on the rEGO support simply by using different metal salts precursor solutions. The RuO$_2$-rEGO composite with an average RuO$_2$ particle size of 2.8 nm shows an extraordinary activity for OER. The overpotential required to reach the current density of 10 mA cm$^{-2}$ is as small as 225 mV for the RuO$_2$-rEGO-250HT catalyst, outperforming the majority of Ru based catalysts reported in the literature. The Pt-rEGO catalyst with a Pt loading of 10.2 wt.% exhibits enhanced performance for HER compared with the CM Pt/C catalyst with a Pt loading of 20 wt.%, leading to more than tripled mass activity. This versatile, simple and scalable method can be easily adapted for the synthesis of various types graphene based functional nanocomposites, for diverse applications beyond electrochemical energy storage and conversions, such as photocatalytic materials, biomedicine and biotechnology.

**Supporting Information**

Supporting Information is available from the Wiley Online Library or from the author.

**Acknowledgements**

Y.P. and J.C. contributed equally to this work. This work made use of the facilities at the University of Manchester Electron Microscopy Center, Y.P. thanks Matthew Smith for the TEM training and technical support. The authors acknowledge the use of the Department of Materials X-ray Diffraction Suite at the University of Manchester and are grateful for the technical support and advice by Dr. John E. Warren. J.Y. thanks The University of Manchester for the award of a Presidential Doctoral Scholarship. The authors acknowledge support for equipment funded via EPSRC (UK) grants to the Sir Henry Royce Institute, grant references EP/S019367/1 and EP/P025021/1. R.A.W.D acknowledges further support from EPSRC (grant reference EP/N032888/1). J.C. and I.A.K. acknowledge the European Union's Horizon 2020 research and innovation program under grant agreement No 785219. I.A.K. also acknowledges the Royal Academy of Engineering and Morgan Advanced Materials for his Chair.

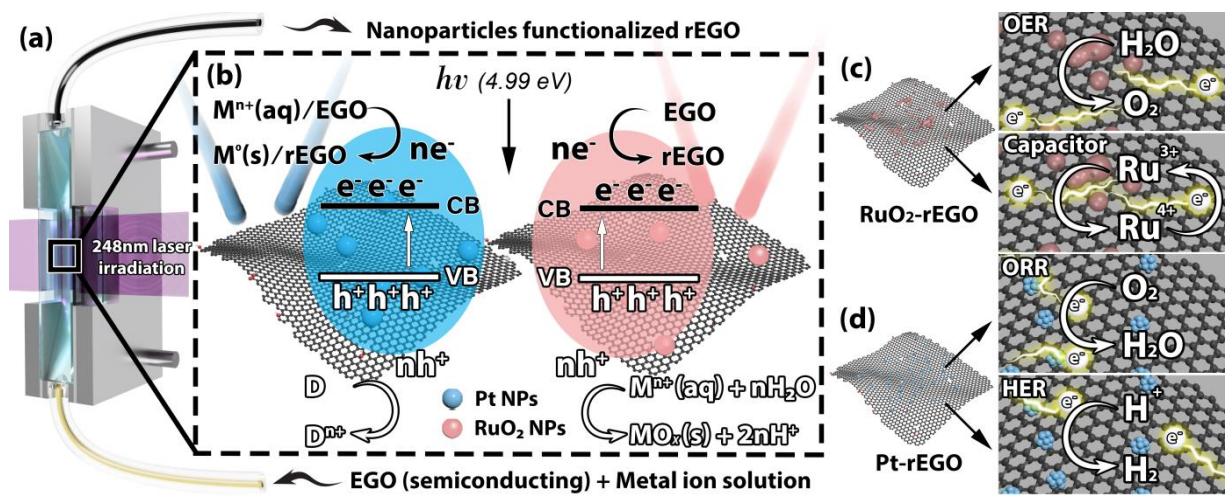

**Figure 1.** Schematic illustration of the laser-assisted synthetic route for rEGO supported nanoparticles. a) The precursor solution of EGO and metal salts (e.g. $RuCl_3$, $H_2PtCl_6$ etc.) is circulated through the quartz cell reactor, which is irradiated by a 248 nm laser beam (photon energy: 4.99 eV; beam size: $1.1 \times 0.4$ cm$^2$). b) The photoexcitation of the semiconducting EGO creates electron-hole pairs, which reduce/oxidize the metal ions ($M^{n+}$) into metal/metal oxide; the metal/metal oxide nucleates and grows on the EGO substrate as nanoparticles; simultaneously, the EGO is reduced by photo-generated electrons to rEGO; Pt (left) and $RuO_2$ (right) nanoparticels (NPs) are shown as expamples for the reductive and oxidative photodeposition, respectively; the photo-generated holes for reductive photodeposition are consumed by sacrifical electron donor (D). c) and d) The as-synthesized graphene supported $RuO_2$ and Pt nanoparticles, respectively, are used for electrochemical energy storage and conversion.

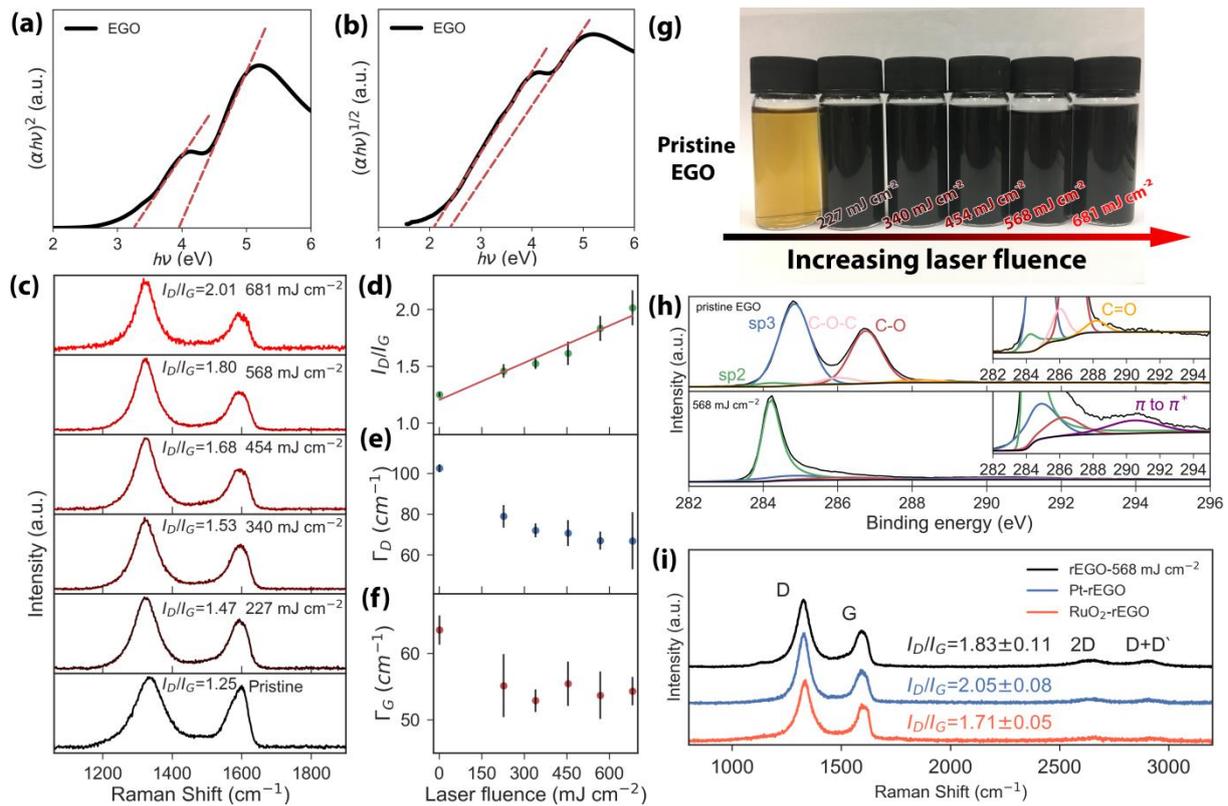

**Figure 2.** a) and b) Tauc plots derived from the UV-vis spectra of pristine EGO for determination of bandgaps for direct and indirect transitions, respectively. c) Typical Raman spectra of EGO after 248 nm laser irradiation at various fluences, d) the evolution of $I_D/I_G$ ratio with the increase of laser fluence. e) and f) FWHM values of D and G band for EGO treated with various laser fluences. g) Photographs of pristine EGO and rEGO after laser irradiation at various fluences. h) XPS high resolution C 1s spectra of pristine EGO and rEGO reduced by laser irradiation at 568 mJ cm$^{-2}$. i) Comparison of typical Raman spectra for pure rEGO, RuO$_2$-rEGO and Pt-rEGO composites prepared using aser irradiation at 568 mJ cm$^{-2}$, the $I_D/I_G$ ratio displayed is an average value of five spectra.

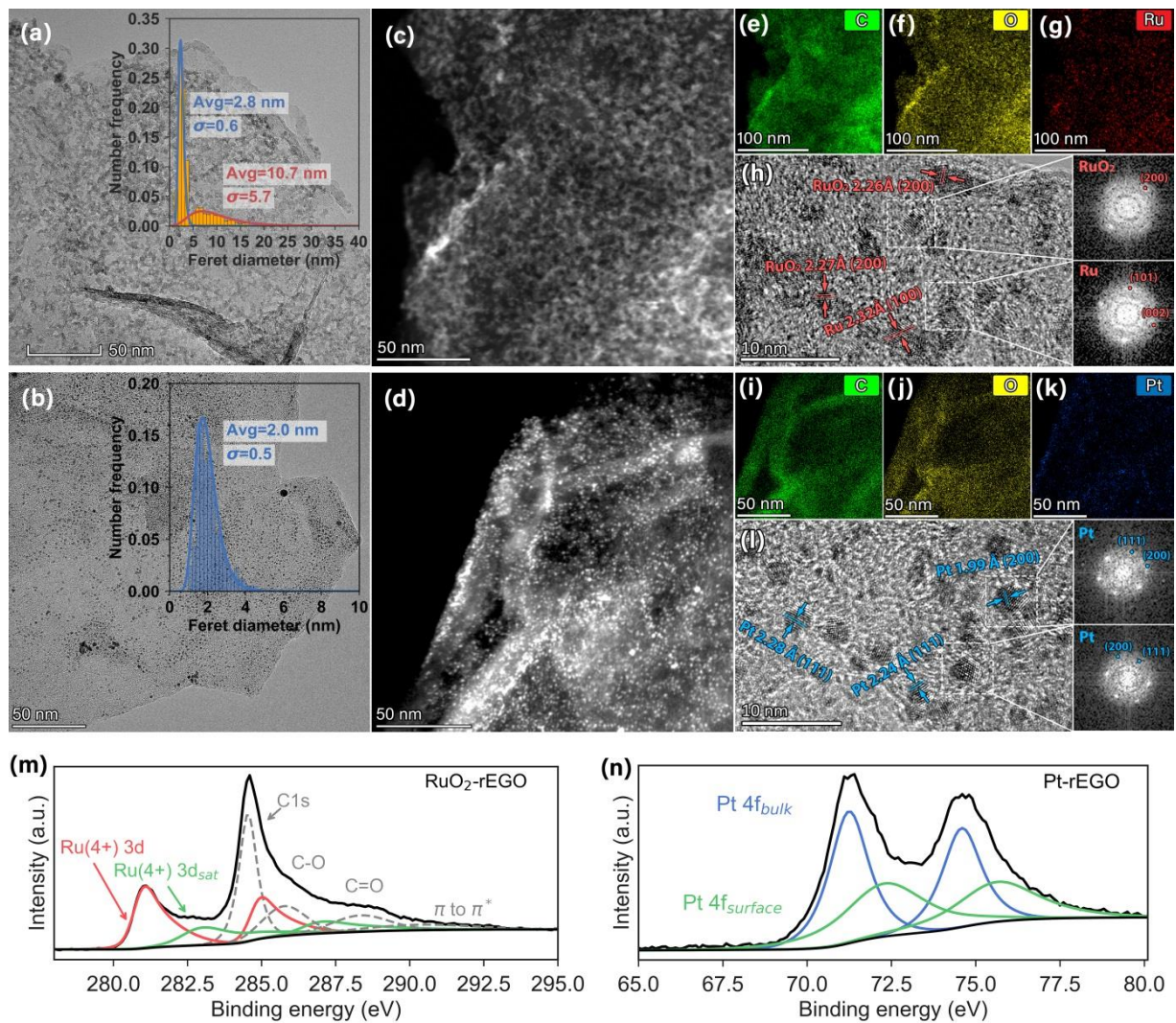

**Figure 3.** TEM images of a) RuO$_2$-rEGO-250HT (heat treated at 250 °C) and b) Pt-rEGO prepared via laser irradiation at 568 mJ cm$^{-2}$, insets: size distribution of two samples by counting 833 and 1000 particles, respectively. STEM-HAADF images of c) RuO$_2$-rEGO-250HT and d) Pt-rEGO. e-g) STEM-EDS elemental mappings of C, O and Ru in the RuO$_2$-rEGO-250HT sample. h) HRTEM image of RuO$_2$-rEGO-250HT shows lattice fringes of Ru and RuO$_2$, inset: representative fast Fourier transform (FFT) pattern of selected particles. i-k) STEM-EDS elemental mappings of C, O and Pt in the Pt-rEGO. l) HRTEM image of Pt-rEGO shows lattice fringes of Pt, inset: representative FFT pattern of selected particles. m) and n) High resolution XPS Ru 3d and Pt 4f spectra recorded from the as-synthesized RuO$_2$-rEGO and Pt-rEGO samples, respectively.

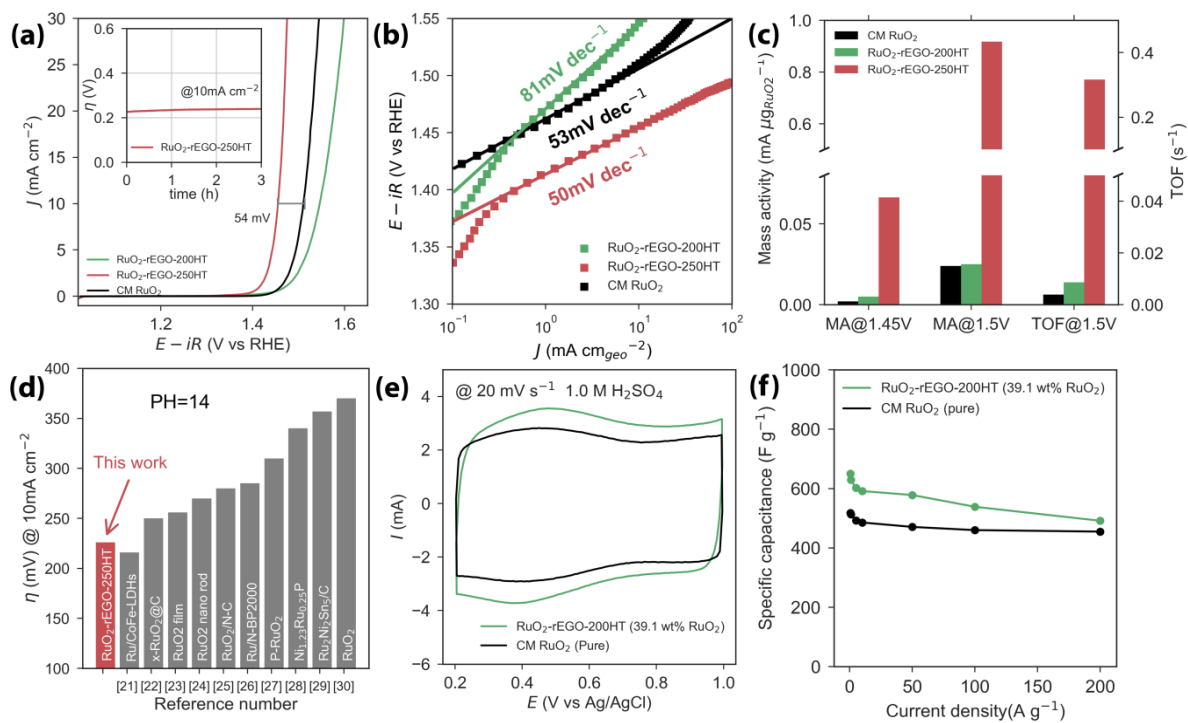

**Figure 4.** a) LSVs of RuO$_2$-rEGO-200HT, RuO$_2$-rEGO-250HT and CM RuO$_2$ measured at a scan rate of 10 mV s$^{-1}$ in 1.0 M KOH electrolyte at 2000 rpm (inset: chronopotentiometry of RuO$_2$-rEGO-250HT catalyst at a current density of 10 mA cm$^{-2}$). b) Tafel plots derived from LSV curves. c) Comparison of mass activities at current densities of 1.45 V and 1.5 V, respectively, the comparison of TOF values at 1.5 V is also included. d) Comparison of the overpotentials required to achieved a 10 mA cm$^{-2}$ current density for various types of literature reported Ru based electrocatalysts. Note: the bibliographic information of the reference numbers can be found in **Table S1** in Supporting Information. e) CVs recorded at 20 mV s$^{-1}$ and f) specific capacitances at various current densities for the RuO$_2$-rEGO-200HT and CM RuO$_2$ electrodes in 1 M H$_2$SO$_4$ aqueous electrolyte.

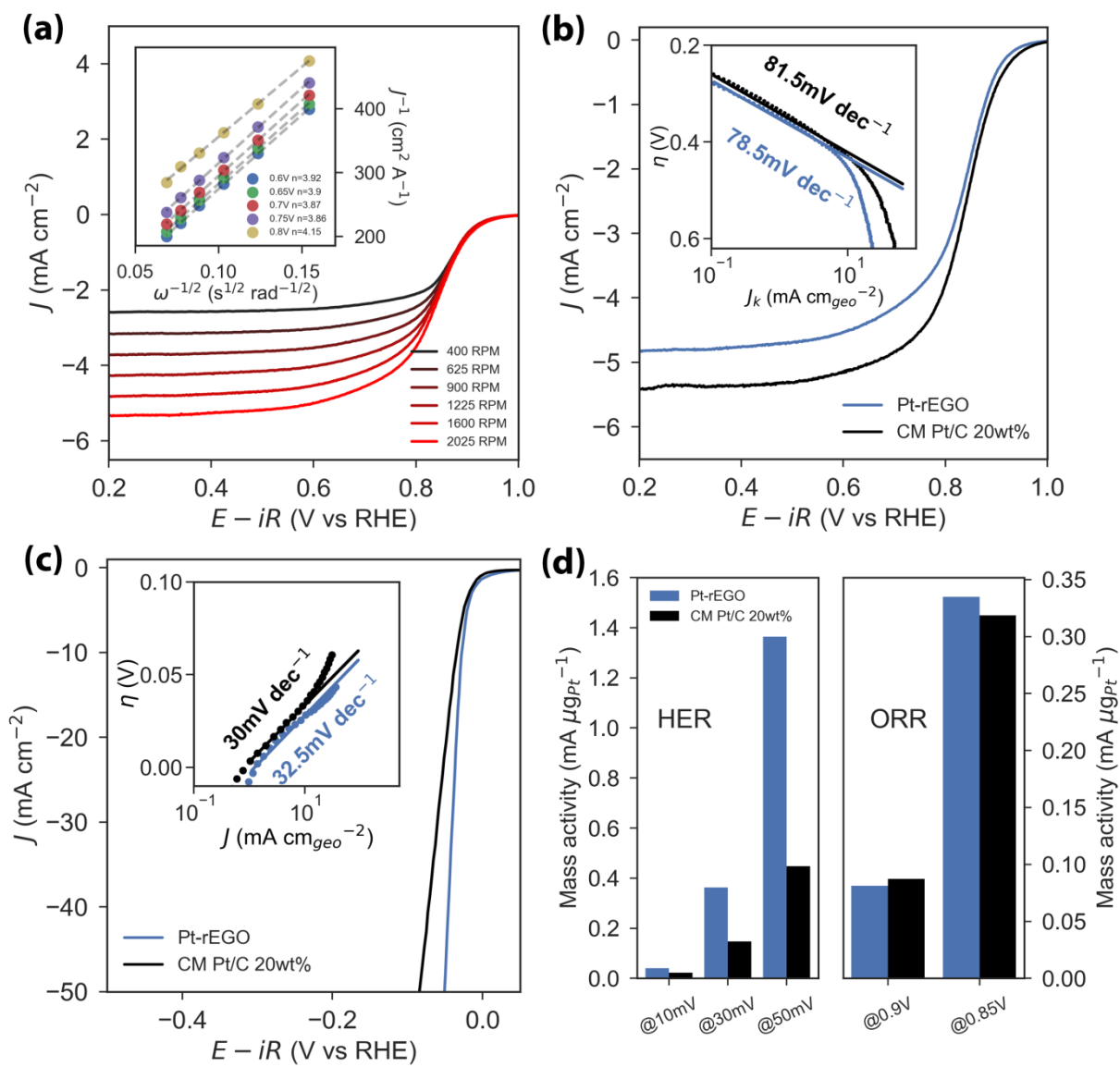

**Figure 5.** a) LSVs of Pt-rEGO tested at different rotation speed from 400 to 2025 rpm at a scan rate of 10 mV s$^{-1}$ in O$_2$-saturated 0.1 M KOH electrolyte, the inset shows the corresponding Koutechy-Levich plots. b) Comparison of LSVs for Pt-rEGO and 20 wt.% Pt/C at 1600 rpm, the inset shows the corresponding Tafel plots. c) LSVs of Pt-rEGO and CM Pt/C 20 wt.% in N$_2$-saturated 0.5 M H$_2$SO$_4$ electrolyte at a scan rate of 10 mV s$^{-1}$ and a rotation speed of 2000 rpm, and the inset image shows the corresponding Tafel plots. d) Comparison of mass activities of Pt-rEGO and CM Pt/C 20 wt.% catalysts for both HER (left panel) and OER (right panel) at different overpotentials.

**Graphene supported electrocatalysts**, including $RuO_2$ and Pt, have been synthesized by laser (248 nm) irradiation of semiconducting electrochemical graphene oxide (EGO) and metal salts precursor solutions. The catalysts show superior electrocatalytic activities for oxygen evolution reaction (OER) and hydrogen evolution reaction (HER), which is attributed to the homogeneous distribution of ultrafine nanoparticles (~2 nm) on the deeply reduced EGO support.

**Electrocatalysts**

Y. Peng, J. Cao*, J. Yang, W. Yang, C. Zhang, X. Li, R. A. W. Dryfe, L. Li, I. A. Kinloch*, Z. Liu*

**Laser Assisted Solution Synthesis of High Performance Graphene Supported Electrocatalysts**

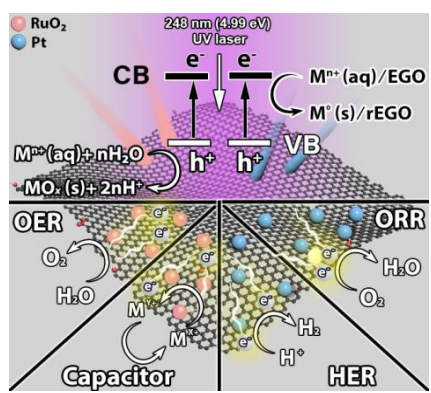